\documentclass[aps,prd,a4paper,noshowpacs]{revtex4}
\begin{document}
\title{Aspects of spin-dependent dark matter search}	
\author{V.A.~Bednyakov}
\affiliation{Laboratory of Nuclear Problems,
         Joint Institute for Nuclear Research, \\ 
         141980 Dubna, Russia; E-mail: Vadim.Bednyakov@jinr.ru}

\begin{abstract} 
	The Weakly Interacting Massive Particle (WIMP)
	is the main candidate for the relic dark matter.  	
	A set of exclusion curves currently obtained 
	for cross sections of the 
	spin-dependent WIMP-proton and WIMP-neutron
	interaction is given.
	A two-orders-of-magnitude improvement 
	of the sensitivity of the 
	dark matter experiment is needed 
	to reach the SUSY predictions for relic neutralinos.
	It is noted that near-future experiments with the high-spin isotope 
	$^{73}$Ge can yield a new important constraint 
	on the neutralino-neutron effective coupling and 
	the SUSY parameter space.
\end{abstract} 

\maketitle 

\subsection{Introduction}
\begin{figure}[t!] 
\begin{picture}(60,140) 
\put(-63,-63){\includegraphics{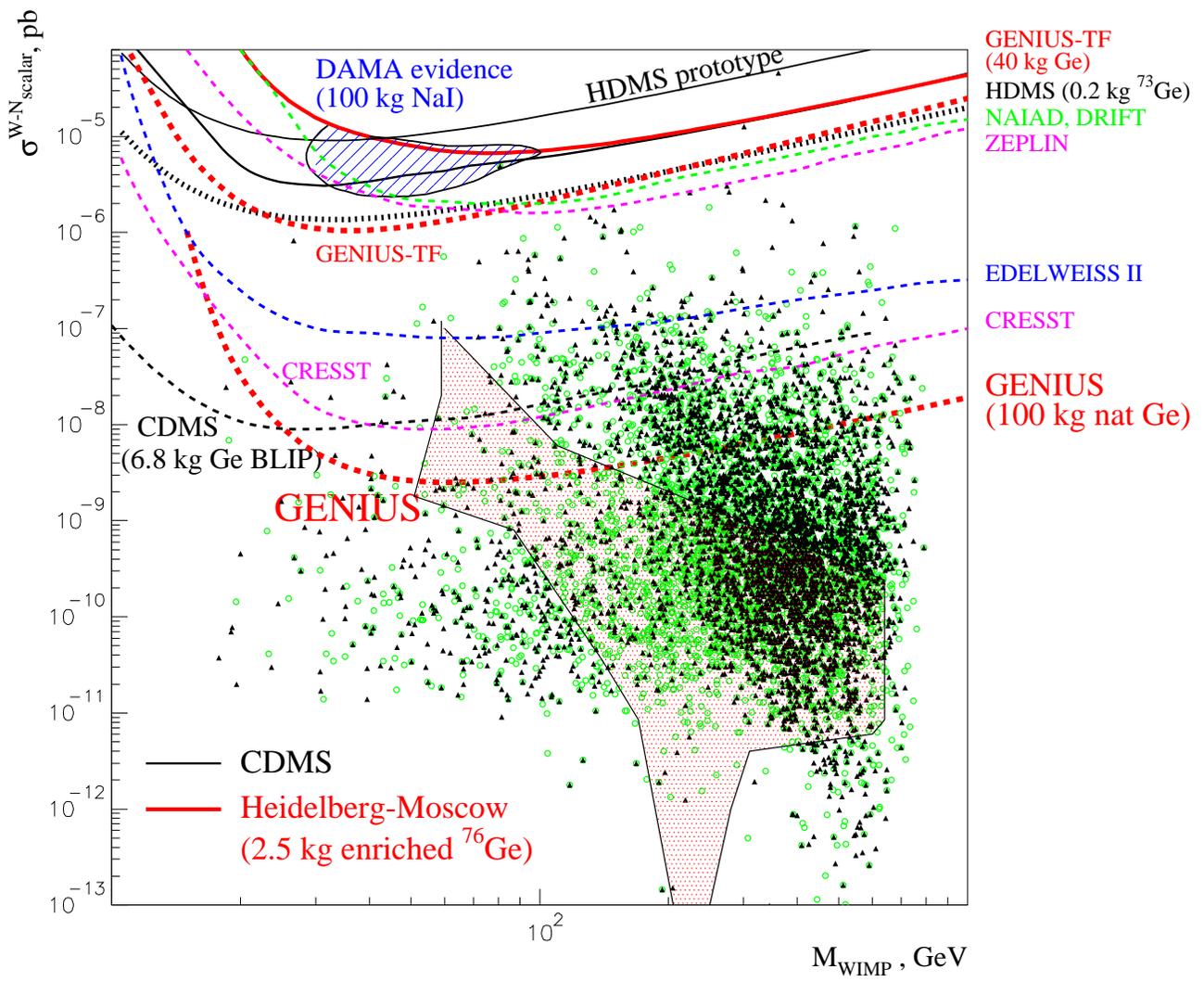}}
\end{picture}
\caption{WIMP-nucleon cross section 
	limits in pb for scalar (spin-independent) interactions as 
        a function of the WIMP mass in GeV. 
        Shown are contour lines of the present experimental limits 
	(solid lines) and of projected experiments (dashed lines). 
        Also shown is the region of evidence published by DAMA. 
        The theoretical expectations are shown by  scatter plots 
	(circles and triangles are from
\cite{Bednyakov:2000he,Bednyakov:2002dz}) 
	and by grey closed region 
\cite{Ellis:2000ds}.
	From 
\cite{Krivosheina:2002pv}. }
\label{Scalar-2002} 
\end{figure}

	Nowadays the main efforts in the direct dark matter search 
	experiments are concentrated in the field of so-called 
	spin-independent (or scalar) interaction of a dark matter particle 
	or the Weakly Interacting Massive Particle (WIMP), with a nucleus. 
	The lightest supersymmetric (SUSY) 
	partilce (LSP) neutralino is
	assumed here as a best WIMP candidate.	
	It is believed that this spin-independent (SI) interaction of 
	dark matter (DM) particles with nuclei makes a dominant 
	contribution to the expected event rate of detection of 
	these particles. 
	The reason is the strong (proportional to the squared mass 
	of the target nucleus) 
	enhancement of SI WIMP-nucleus interaction.
	The results currently obtained in the field are
	usually presented in the form of exclusion curves
(see for example Fig.~\ref{Scalar-2002}).
	For the fixed mass of the WIMP the values of the cross section
	due to scalar elastic WIMP-nucleon interaction  
	located above these curves are already excluded experimentally.  
	There is also the DAMA closed contour which corresponds to
	the first claim for evidence for the dark matter signal
	due to the positive annual modulation effect
\cite{Bernabei:2000qi}.   

	In the paper we consider some aspects of the spin-dependent 
	(or axial-vector) interaction of the DM WIMP with nuclei.
	There are at least three reasons to think that this 
	spin-dependent (SD) interaction could also be very important. 
	First, contrary to the only one constraint for SUSY models available 
	from scalar WIMP-nucleus interaction, the spin WIMP-nucleus 
	interaction supplies us with two such constraints 
(see for example 
\cite{Bednyakov:1994te} and formulas below).
	Second, one can notice 
\cite{Bednyakov:2000he,Bednyakov:2002mb}
	that even with a very accurate DM detector
	(say, with sensitivity $10^{-5}\,$events$/$day$/$kg)
	which is sensitive only to the WIMP-nucleus 
	scalar interaction (with spinless target nuclei) 
	one can, in principle, miss a DM signal. 
	To safely avoid such a situation one should
	have a spin-sensitive DM detector, i.e. a detector 
	with spin-non-zero target nuclei.
	Finally, there is a complicated (and theoretically very interesting) 
	nucleus spin structure, which possesses 
	the so-called long $q$-tail form-factor behavior 
	for heavy targets and heavy WIMP.
	Therefore, the SD efficiency to detect a DM signal 
	is much higher than the SI efficiency,
	especially for the heavy target nucleus and WIMP masses
\cite{Engel:1991wq}.

\subsection{Zero Momentum Transfer}
        A dark matter event is elastic scattering 
	of a relic neutralino $\chi$ (or $\tilde\chi$)\ 
	from a target nucleus $A$ producing a nuclear 
	recoil $E_{\rm R}$ which can be detected by a suitable detector.
	The differential event rate in respect to the recoil 
	energy is the subject of experimental measurements.
	The rate depends on the distribution of
        the relic neutralinos in the solar vicinity $f(v)$ and
        the cross section of neutralino-nucleus elastic scattering
\cite{Jungman:1996df,Lewin:1996rx,Smith:1990kw,Bednyakov:1999yr,Bednyakov:1996yt,Bednyakov:1997ax,Bednyakov:1997jr,Bednyakov:1994qa}.
	The differential event rate per unit mass of 
	the target material has the form
\begin{equation}
\label{Definitions.diff.rate}
	\frac{dR}{dE_{\rm R}} = N \frac{\rho_\chi}{m_\chi}
	\displaystyle
	\int^{v_{\rm max}}_{v_{\rm min}} dv f(v) v
	{\frac{d\sigma}{dq^2}} (v, q^2). 
\end{equation}
        The nuclear recoil energy
	$E_{\rm R} = q^2 /(2 M_A )$ is typically about $10^{-6} m_{\chi}$ and 
	$N={\cal N}/A$ is the number density of target nuclei, where 
	${\cal N}$ is the Avogadro number and $A$ is the atomic mass 
	of the nuclei with mass $M_A$. 
	The neutralino-nucleus elastic scattering cross section 
	for spin-non-zero ($J\neq 0$) 
	nuclei contains coherent (spin-independent, or
	SI) and axial (spin-dependent, or SD) terms
\cite{Engel:1992bf,Engel:1991wq,Ressell:1993qm}: 
\begin{eqnarray}
\nonumber
{\frac{d\sigma^{A}}{dq^2}}(v,q^2) 
	&=& \frac{\sum{|{\cal M}|^2}}{\pi\, v^2 (2J+1)} 
         =  \frac{S^A_{\rm SD} (q^2)}{v^2 (2J+1)} 
           +\frac{S^A_{\rm SI} (q^2)}{v^2 (2J+1)} \\
\label{Definitions.cross.section}
        &=& \frac{\sigma^A_{\rm SD}(0)}{4\mu_A^2 v^2}F^2_{\rm SD}(q^2)
           +\frac{\sigma^A_{\rm SI}(0)}{4\mu_A^2 v^2}F^2_{\rm SI}(q^2).
\end{eqnarray} 
	The normalized non-zero-momentum-transfer nuclear form-factors
\begin{equation}
\label{Definitions.form.factors}
F^2_{\rm SD,SI}(q^2) = \frac{S^{A}_{\rm SD,SI}(q^2)}{S^{A}_{\rm SD,SI}(0)}
\qquad (F^2_{\rm SD,SI}(0) = 1),
\end{equation}
	are defined via nuclear structure functions 
\cite{Engel:1992bf,Engel:1991wq,Ressell:1993qm} 
\begin{eqnarray}
\label{Definitions.scalar.structure.function}
S^{A}_{\rm SI}(q) 
	&=& 
	\sum_{L\, {\rm even}} 
        \vert\langle J \vert\vert {\cal C}_L(q) \vert\vert J \rangle \vert^2 
	\simeq  
	\vert\langle J \vert\vert {\cal C}_0(q) \vert\vert J \rangle \vert^2 ,
\\[3pt]
\label{Definitions.spin.structure.function}
S^A_{\rm SD}(q) 
	&=& 
	\sum_{L\, {\rm odd}} \big( 
	\vert\langle N \vert\vert {\cal T}^{el5}_L(q) 
	\vert\vert N \rangle\vert^2 + \vert\langle N \vert\vert 
	{\cal L}^5_L (q) \vert\vert N \rangle\vert^2\big). 
\end{eqnarray} 
	The transverse electric ${\cal T}^{el5}(q)$ 
	and longitudinal ${\cal L}^5(q)$ multipole projections of the
	axial vector current operator, scalar function ${\cal C}_L(q)$ 
	are given in the form
\begin{eqnarray*}
{\cal T}^{el5}_L(q) 
        &=& \frac{1}{\sqrt{2L+1}}\sum_i\frac{a^{}_0 +a^{}_1\tau^i_3}{2}
                 \Bigl[
		-\sqrt{L}   M_{L,L+1}(q\vec{r}_i)
                +\sqrt{L+1} M_{L,L-1}(q\vec{r}_i)
                 \Bigr], \\
 {\cal L}^5_L(q)
         &=& \frac{1}{\sqrt{2L+1}}\sum_i
                  \Bigl( \frac{a^{}_0}{2} +
                     \frac{a^{}_1 m^2_\pi \tau^i_3}{2(q^2+m_\pi^2)}
                  \Bigr)                 
                  \Bigl[
                 \sqrt{L+1} M_{L,L+1}(q\vec{r}_i)
		+\sqrt{L}   M_{L,L-1}(q\vec{r}_i)
                 \Bigr], \\[5pt]
{\cal C}_L(q) 
	&=& \sum_{i,\ {\rm nucleons}} c_0^{} j_L(qr_i)Y_L(\hat{r}_i), \  \  \ 
{\cal C}_0(q) = \sum_{i} c_0^{} j_0(qr_i)Y_0(\hat{r}_i),  
\end{eqnarray*} 
	where $a^{}_{0,1} = a^{}_n \pm a^{}_p$ 
(see (\ref{Definitions.spin.zero.cs})) and
$M_{L,L'}(q\vec{r}_i) = j_{L'}(qr_i)[Y_{L'}(\hat{r}_i)\vec{\sigma}_i]^L$
\cite{Engel:1992bf,Engel:1991wq,Ressell:1993qm}.
	The nuclear SD and SI cross sections at $q=0$ 
(in (\ref{Definitions.cross.section})) have the forms  
\begin{eqnarray}
\label{Definitions.scalar.zero.momentum}
\sigma^A_{\rm SI}(0) 
	&=& \frac{4\mu_A^2 \ S^{}_{\rm SI}(0)}{(2J+1)}\! =\!
	     \frac{\mu_A^2}{\mu^2_p}A^2 \sigma^{p}_{{\rm SI}}(0), \\ 
\label{Definitions.spin.zero.momentum}
\sigma^A_{\rm SD}(0)
	&=&  \frac{4\mu_A^2 S^{}_{\rm SD}(0)}{(2J+1)}\! =\!
	     \frac{4\mu_A^2}{\pi}\frac{(J+1)}{J}
             \left\{a_p\langle {\bf S}^A_p\rangle 
                  + a_n\langle {\bf S}^A_n\rangle\right\}^2\\
\label{Definitions.spin.zero.momentum.Tovei}
      &=&
	\frac{\mu_A^2}{\mu^2_{p,n}}\frac{(J+1)}{3\, J}
\left\{ \sqrt{\sigma^{p}_{{\rm SD}}(0)}\langle{\bf S}^A_p\rangle 
       +{\rm sign}(a_p\, a_n)
	\sqrt{\sigma^{n}_{{\rm SD}}(0)}\langle{\bf S}^A_n\rangle\right\}^2.
\end{eqnarray}
	Here $\displaystyle \mu_A = \frac{m_\chi M_A}{m_\chi+ M_A}$
	is the reduced neutralino-nucleus mass.
	The zero-momentum-transfer 
	proton and neutron SI and SD cross sections
\begin{eqnarray}
\label{Definitions.scalar.zero.cs}
\sigma^{p}_{{\rm SI}}(0) 
	= 4 \frac{\mu_p^2}{\pi}c_{0}^2,
&\qquad&
	c^{}_{0} \equiv	c^{(p,n)}_0 = \sum_q {\cal C}_{q} f^{(p,n)}_q; \\
\label{Definitions.spin.zero.cs}
\sigma^{p,n}_{{\rm SD}}(0)  
 	=  12 \frac{\mu_{p,n}^2}{\pi}{a}^2_{p,n} 
&\qquad&
	a_n =\sum_q {\cal A}_{q} \Delta^{(p)}_q, \quad 
	a_p =\sum_q {\cal A}_{q} \Delta^{(n)}_q
\end{eqnarray}
	depend on the effective neutralino-quark scalar 
	${\cal C}_{q}$ and axial-vector ${\cal A}_{q}$ couplings 
	from the effective Lagrangian
\begin{equation}
\label{Definitions.effective.lagrangian}
{\cal  L}_{\rm eff} = \sum_{q}^{}\left( 
	{\cal A}_{q}\cdot
      \bar\chi\gamma_\mu\gamma_5\chi\cdot
                \bar q\gamma^\mu\gamma_5 q + 
	{\cal C}_{q}\cdot\bar\chi\chi\cdot\bar q q
	\right)
      \ + ... 
\end{equation}
	and on the spin ($\Delta^{(p,n)}_q$)
	and mass ($f^{(p,n)}_q$) structure of nucleons.
	The factors $\Delta_{q}^{(p,n)}$ parametrize the quark 
	spin content of the nucleon and are defined by the relation
	$ \displaystyle 2 \Delta_q^{(n,p)} s^\mu  \equiv 
          \langle p,s| \bar{\psi}_q\gamma^\mu \gamma_5 \psi_q    
          |p,s \rangle_{(p,n)}$.
	The total nuclear spin (proton, neutron) operator 
	is defined as follows 
\begin{equation}
\label{Definitions.spin.operator}
 {\bf S}_{p,n} = \sum_i^A {\bf s}_{p,n} ({i}),
\end{equation}
	where $i$ runs over all nucleons.
	Further the convention is used 
	that all angular momentum operators 
	are evaluated in their $z$-projection 
	in the maximal $M_J$ state, e.g.
\begin{equation}
\label{Definitions.spin.operator.1}
\langle {\bf S} \rangle \equiv \langle N \vert {\bf S} \vert N \rangle
\equiv  \langle J,M_J = J \vert S_z \vert J,M_J = J \rangle.
\end{equation}
	Therefore
	$ \langle {\bf S}_{p(n)} \rangle $ is the spin of the proton 
	(neutron) averaged over all nucleons in the nucleus $A$.
	The cross sections at zero momentum transfer show strong 
	dependence on the nuclear structure of the ground state
\cite{Engel:1995gw,Ressell:1997kx,Divari:2000dc}. 

	The relic neutralinos in the halo of our Galaxy have a mean velocity of
	$\langle v \rangle  \simeq 300~{\rm km/s} = 10^{-3} c$.  
	When the product $q_{\rm max}R \ll 1$, 
	where $R$ is the nuclear radius 
	and $q_{\rm max} = 2 \mu_A v$ is the maximum momentum 
	transfer in the $\tilde\chi A$ scattering, the matrix element 
	for the spin-dependent $\tilde\chi A$ 
	scattering reduces to a very simple form
({\em zero momentum transfer limit})\
\cite{Engel:1995gw,Ressell:1997kx}:
\begin{equation}
\label{Definitions.matrix.element}
 {\cal M} = C \langle N\vert a_p {\bf S}_p + a_n {\bf S}_n
 	\vert N \rangle \cdot {\bf s}_{\tilde \chi}
 	  = C \Lambda \langle N\vert {\bf J}
	 \vert N \rangle \cdot {\bf s}_{\tilde \chi}.
\end{equation}
	Here ${\bf s}_{\chi}$ is the spin of the neutralino, and 
\begin{equation}
 \Lambda = {{\langle N\vert a_p {\bf S}_p + a_n {\bf S}_n
\vert N \rangle}\over{\langle N\vert {\bf J}
\vert N \rangle}} =
{{\langle N\vert ( a_p {\bf S}_p + a_n {\bf S}_n ) \cdot {\bf J}
\vert N \rangle}\over{ J(J+1)
}}. 
\end{equation}
	It is seen that the $\chi$ couples to the spin carried
	by the protons and the neutrons.  
	The normalization $C$ involves the coupling
	constants, masses of the exchanged bosons and various LSP
	mixing parameters that have no effect upon the nuclear matrix element
\cite{Griest:1988ma}.  
	In the limit of zero momentum transfer $q=0$ 
	the spin structure function 
(\ref{Definitions.spin.structure.function}) reduces to
\begin{equation}
S^A_{\rm SD}(0) = 
	{2 J + 1\over{\pi}} \Lambda^2 J(J + 1). 
\end{equation}

	Perhaps the first model to estimate the spin content in the nucleus
	for the dark matter search was the 
	independent single-particle shell model ({\bf ISPSM})
	used originally by Goodman and Witten 
\cite{Goodman:1985dc} and later in
\cite{Drukier:1986tm,Ellis:1988sh,Smith:1990kw}.
	The ground state value of the nuclear total spin $J$ can be 
	described by those of one extra nucleon interacting with the
	effective potential of the nuclear core.  

	There are nuclear structure calculations 
	(including non-zero-momentum approximation) for spin-dependent
	neutralino interaction with 
	helium $^3$He 
\cite{Vergados:1996hs};
	fluorine $^{19}$F
\cite{Vergados:2002bb,Divari:2000dc,Vergados:1996hs};
	sodium $^{23}$Na
\cite{Vergados:2002bb,Ressell:1997kx,Divari:2000dc,Vergados:1996hs};
	aluminium $^{27}$Al
\cite{Engel:1995gw};
	silicon $^{29}$Si
\cite{Vergados:2002bb,Ressell:1993qm,Divari:2000dc};
	chlorine $^{35}$Cl
\cite{Ressell:1993qm};
	potassium $^{39}$K
\cite{Engel:1995gw};
	germanium $^{73}$Ge 
\cite{Ressell:1993qm,Dimitrov:1995gc};
	niobium $^{93}$Nd
\cite{Engel:1992qb};
	iodine $^{127}$I
\cite{Ressell:1997kx};
	xenon $^{129}$Xe
\cite{Ressell:1997kx} and
	$^{131}$Xe
\cite{Ressell:1997kx,Nikolaev:1993vw,Engel:1991wq}; 
	tellurium $^{123}$Te
\cite{Nikolaev:1993vw}
	and 
	$^{125}$Te
\cite{Ressell:1997kx};
	lead $^{208}$Pb
\cite{Kosmas:1997jm,Vergados:1996hs}.
	The zero-momentum case 
	is also investigated for 
	Cd, Cs, Ba and La in 
\cite{Pacheco:1989jz,Nikolaev:1993vw,Iachello:1991ut}.

	There are several approaches (advocated by a few groups 
	of researchers) to the more accurate calculation of the nuclear
	structure effects relevant to the dark matter detection.
	To the best of our knowledge an almost 
	full list of the models includes the
	Odd Group Model ({\bf OGM}) of Engel and Vogel
\cite{Engel:1989ix} and their extended OGM ({\bf EOGM})
\cite{Engel:1989ix,Engel:1992bf}; 
	Interacting Boson Fermion Model ({\bf IBFM}) of
	Iachello, Krauss, and Maino 	 
\cite{Iachello:1991ut};
	Theory of Finite Fermi Systems ({\bf TFFS}) of 
	Nikolaev and Klapdor-Kleingrothaus
\cite{Nikolaev:1993dd};
	Quasi Tamm-Dancoff Approximation ({\bf QTDA}) of Engel
\cite{Engel:1991wq};
	different shell model treatments ({\bf SM}) by Pacheco and Strottman 
\cite{Pacheco:1989jz};
	by Engel, Pittel, Ormand and Vogel 
\cite{Engel:1992qb} and 
	Engel, Ressell, Towner and Ormand,
\cite{Engel:1995gw}, 	
	by Ressell et al.
\cite{Ressell:1993qm} and 
	Ressell and Dean
\cite{Ressell:1997kx};
	by Kosmas, Vergados et al.
\cite{Vergados:1996hs,Kosmas:1997jm,Divari:2000dc};
	so-called ``{\bf hybrid}'' model of Dimitrov, Engel and Pittel 
\cite{Dimitrov:1995gc}
	and perturbation theory ({\bf PT}) based on 
	calculations of Engel et al.
\cite{Engel:1995gw}.

\subsection{Spin constraints}
	For the spin-zero nuclear target the experimentally measured 
	event rate 
(\ref{Definitions.diff.rate}) 
	of direct DM particle detection via formula 
(\ref{Definitions.cross.section}) 
	is connected with zero-momentum 
	WIMP-proton (for the neutron the cross section is the same)
	cross section 
(\ref{Definitions.spin.zero.momentum}). 
	The zero momentum scalar WIMP-proton (neutron) cross section 
	$\sigma^{p}_{{\rm SI}}(0)$ 
	can be expressed through 
	effective neutralino-quark couplings ${\cal C}_{q}$
(\ref{Definitions.effective.lagrangian})	 
	by means of expression 
(\ref{Definitions.scalar.zero.cs}).
	These couplings ${\cal C}_{q}$ (as well as ${\cal A}_{q}$) 
	can be directly connected with the
	fundamental parameters of a SUSY model 
	such as $\tan \beta$, $M_{1,2}$, $\mu$, masses 
	of sfermions and Higgs bosons, etc. 
	Therefore experimental limitations on the 
	spin-independent neutralino-nucleon cross section
	supply us with a constraint on the fundamental parameters
	of an underlying SUSY model.

\begin{figure}[!h] 
\begin{picture}(100,135)
\put(-40,-60){\includegraphics{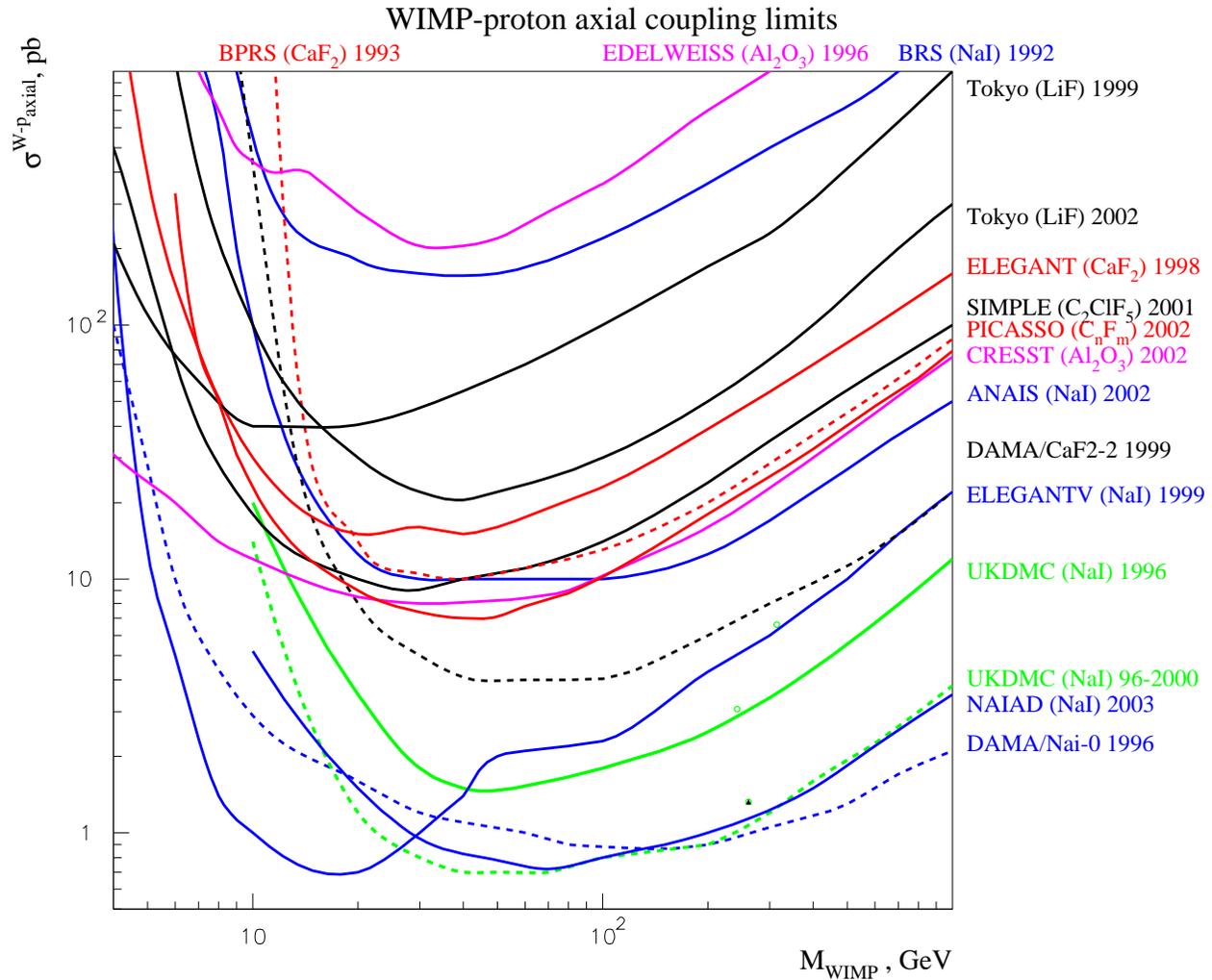}}
\end{picture}
\caption{Full set of currently available exclusion curves for
	spin-dependent WIMP-proton cross sections
	($\sigma^{p}_{{\rm SD}}$ as a function of WIMP mass).
	The curves are obtained from   
\protect\cite{Bacci:1994ip,Bacci:1992pd,Bacci:1994tt,deBellefon:1996jh,%
Bernabei:1996vj,Bernabei:1997qb,Belli:1999xh,%
Sarsa:1996pa,Smith:1996fu,Sumner:1999yw,Spooner:2000kt,Ogawa:2000vi,%
Fushimi:1999kp,Yoshida:2000df,%
Ootani:1998jy,Minowa:1998ai,Ootani:1999pt,Ootani:1999xv,Miuchi:2002zp,%
Collar:2000zw,Angloher:2002in,Boukhira:2002qj,Cebrian:2002vd,Ahmed:2003su}.
}
\label{Spin-p} 
\end{figure} 

\begin{figure}[!h] 
\begin{picture}(100,135)
\put(-40,-60){\includegraphics{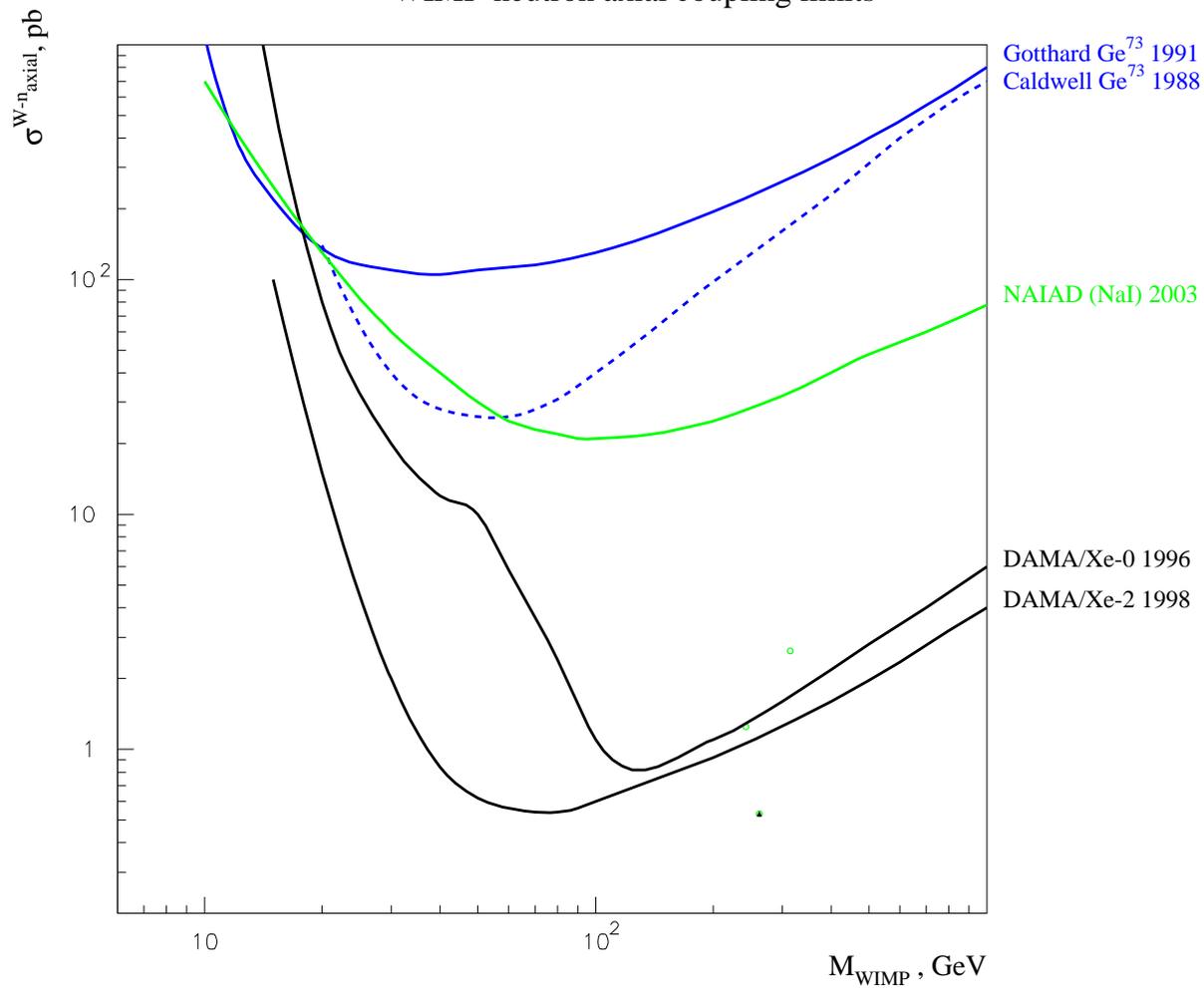}}
\end{picture}
\caption{Full set of currently available exclusion curves for
	spin-dependent WIMP-neutron cross sections
	($\sigma^{n}_{{\rm SD}}$ versus WIMP mass).
	The curves are obtained from 
\protect\cite{Caldwell:1988su,Reusser:1991ri,%
Belli:1996sh,Belli:1996yf,Bernabei:1998ad,Ahmed:2003su}
	Note that the NAIAD curve corresponds to the sub-dominant
	contribution, extracted from the p-odd nucleus Na 
	(compare with the relevant NAIAD curve in 
Fig.~\ref{Spin-p})
}
\label{Spin-n} 
\end{figure} 

	In the case of the spin-dependent WIMP-nucleus interaction
	from measured differential rate 
(\ref{Definitions.diff.rate}) one first extracts 
	limitation for $\sigma^{A}_{{\rm SD}}(0)$ 
	and therefore has in principle two constraints
\cite{Bednyakov:1994te}
	for the neutralino-proton $a^{}_p$ and  
	neutralino-neutron $a^{}_n$ spin effective couplings
	as follows from relation 
(\ref{Definitions.spin.zero.momentum}).
	From 
(\ref{Definitions.spin.zero.momentum})
	one can also see that contrary to spin-independent case
(\ref{Definitions.scalar.zero.momentum})
	there is no factorization of the nuclear structure
	for $\sigma^A_{\rm SD}(0)$.
	Both proton $\langle{\bf S}^A_{p}\rangle$
	and neutron $\langle{\bf S}^A_{n}\rangle$
	spin contributions simultaneously entering formula 
(\ref{Definitions.spin.zero.momentum})
	for the SD WIMP-nuclear
	cross section $\sigma^A_{\rm SD}(0)$.

	In the earlier considerations based on the OGM 
\cite{Engel:1989ix,Engel:1992bf} 
	one assumed that the nuclear spin is carried by the ``odd''
	unpaired group of protons or neutrons and only one of either 
	$\langle{\bf S}^A_n\rangle$ or $\langle{\bf S}^A_p\rangle$ 
	is non-zero (the same is true in the ISPSM 
\cite{Goodman:1985dc,Drukier:1986tm,Ellis:1988sh,Smith:1990kw}).
 	In this case all possible target nuclei can naturally 
	be classified into n-odd and p-odd groups.
	The current experimental situation 
	for the spin-dependent WIMP-{\bf proton} cross section is given in 
Fig.~\ref{Spin-p}. 	
	The data are taken from experiments BRS, (NaI, 1992)
\cite{Bacci:1994ip,Bacci:1992pd}, 
	BPRS (CaF$_2$, 1993)
\cite{Bacci:1994tt},
	EDELWEISS (sapphire, 1996)
\cite{deBellefon:1996jh}, 
	DAMA (NAI, 1996)
\cite{Bernabei:1996vj},
	DAMA (CaF$_2$, 1999)
\cite{Bernabei:1997qb,Belli:1999xh},
	UKDMS (NaI, 1996)
\cite{Sarsa:1996pa,Smith:1996fu,Sumner:1999yw,Spooner:2000kt},
	ELEGANT (CaF$_2$, 1998)
\cite{Ogawa:2000vi},
	ELEGANT (NaI, 1999)
\cite{Fushimi:1999kp,Yoshida:2000df},
	Tokio  (LiF, 1999, 2002)
\cite{Ootani:1998jy,Minowa:1998ai,Ootani:1999pt,Ootani:1999xv,Miuchi:2002zp},
	SIMPLE (C$_{2}$ClF$_{5}$, 2001)
\cite{Collar:2000zw},
	CRESST (Al$_2$O$_3$, 2002)
\cite{Angloher:2002in},
	PICASSO (C$_n$F$_m$, 2002)
\cite{Boukhira:2002qj}, 
	ANAIS (NaI, 2002) 
\cite{Cebrian:2002vd} and
	NAIAD (NaI, 2003)
\cite{Ahmed:2003su}.
	The current experimental situation 
	for the spin-dependent WIMP-{\bf neutron} cross section is given in 
Fig.~\ref{Spin-n}.
	The data are taken from the
	first experiments with natural Ge (1988, 1991)
\cite{Caldwell:1988su,Reusser:1991ri},
	xenon (DAMA/Xe-0,2)
\cite{Belli:1996sh,Belli:1996yf,Bernabei:1998ad}
	and sodium iodide (NAIAD)
\cite{Ahmed:2003su}.
	In the future one can also expect 
	exclusion curves for the SD cross section, for example, 
	from the EDELWEISS and CDMS experiments 
	with natural germanium bolometric detectors.

\begin{figure}[!p] 
\begin{picture}(100,210)
\put(-43,-53){\includegraphics{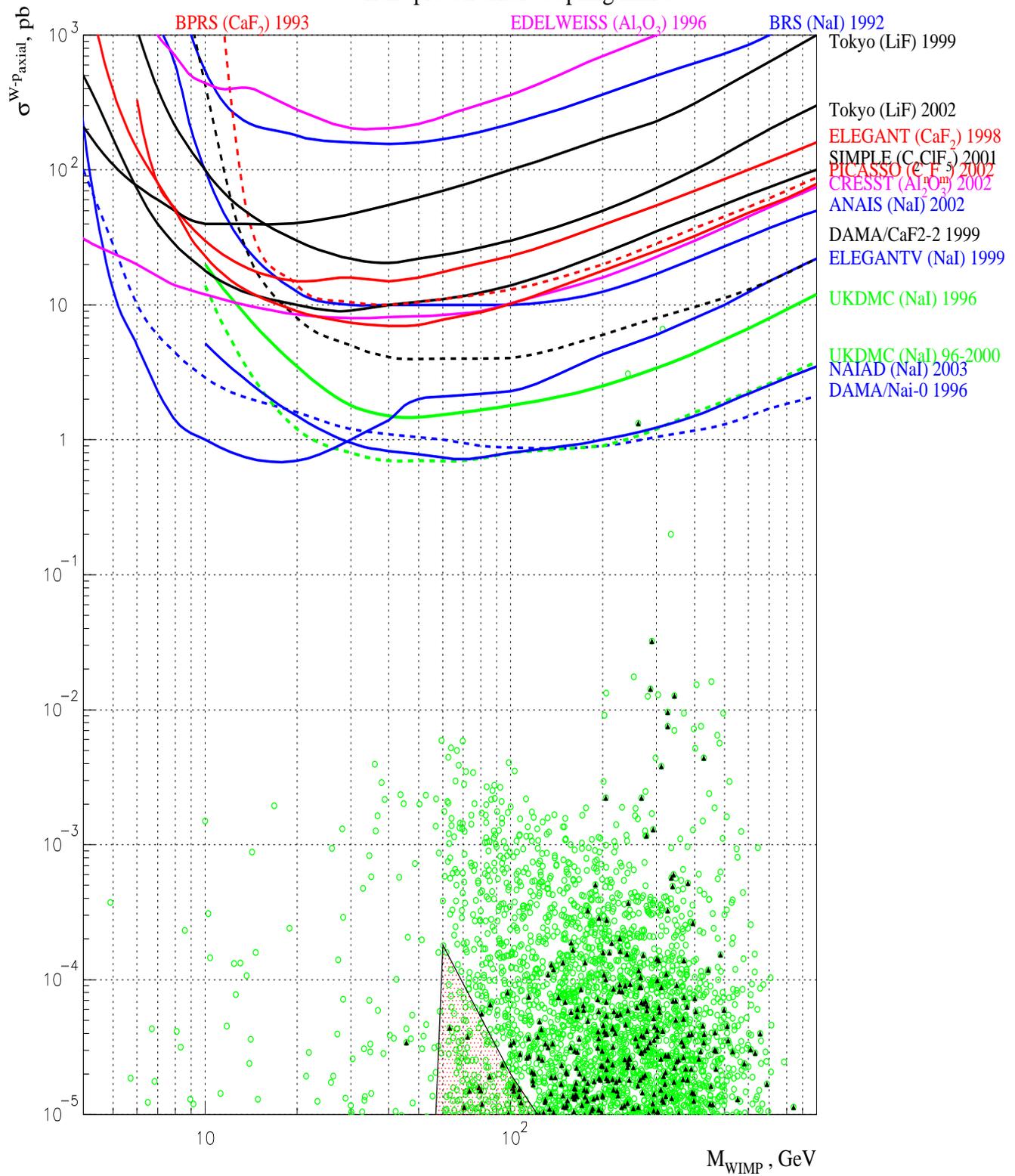}}
\end{picture}
\caption{The same as in 
Fig.~\protect\ref{Spin-p}, but with the theoretical scatter plot from 
\cite{Bednyakov:2002dz}, obtained in the effMSSM with 
	all coannihilation channels included (green circles) and  
	with $~0.1< \Omega h^2 <0.3~$ (black triangles). 
	The triangle-like shaded area is taken from 
\cite{Ellis:2000ds}.
	}
\label{Spin-p-MSSM} 
\end{figure} 
	From 
Fig.~\ref{Spin-p-MSSM} one can conclude that an  
	about two-orders-of-magnitude improvement 
	of the current DM experiment sensitivities is needed 
 	to reach the SUSY predictions for the $\sigma^{p}_{{\rm SD}}$ 
	provided the SUSY lightest neutralino (LSP) 
	is the best WIMP particle candidate. 
	There is the same situation for the $\sigma^{n}_{{\rm SD}}$. 

	Further more accurate calculations of 
\cite{Iachello:1991ut,Engel:1991wq,Pacheco:1989jz,%
      Engel:1992qb,Engel:1995gw,Dimitrov:1995gc,%
      Ressell:1993qm,Ressell:1997kx,%
      Vergados:1996hs,Kosmas:1997jm,Divari:2000dc}
	demonstrate that contrary to the simplified odd-group approach both
	$\langle{\bf S}^A_{p}\rangle$ and $\langle{\bf S}^A_{n}\rangle$ 
	differ from zero, but nevertheless 
	one of these spin quantities always dominates
	($\langle{\bf S}^A_{p}\rangle \ll \langle{\bf S}^A_{n}\rangle$, or
	 $\langle{\bf S}^A_{n}\rangle \ll \langle{\bf S}^A_{p}\rangle$). 
	If together with the dominance like 
	$\langle{\bf S}^A_{p(n)}\rangle \ll \langle{\bf S}^A_{n(p)}\rangle$
	one would have the WIMP-proton and WIMP-neutron couplings
	of the same order of magnitude
	(not $a_{n(p)}\! \ll\! a_{p(n)}$), 
	the situation could look like that in the odd-group model.
\begin{figure}[!h] 
\begin{picture}(100,120)
\put(-20,-44){\includegraphics{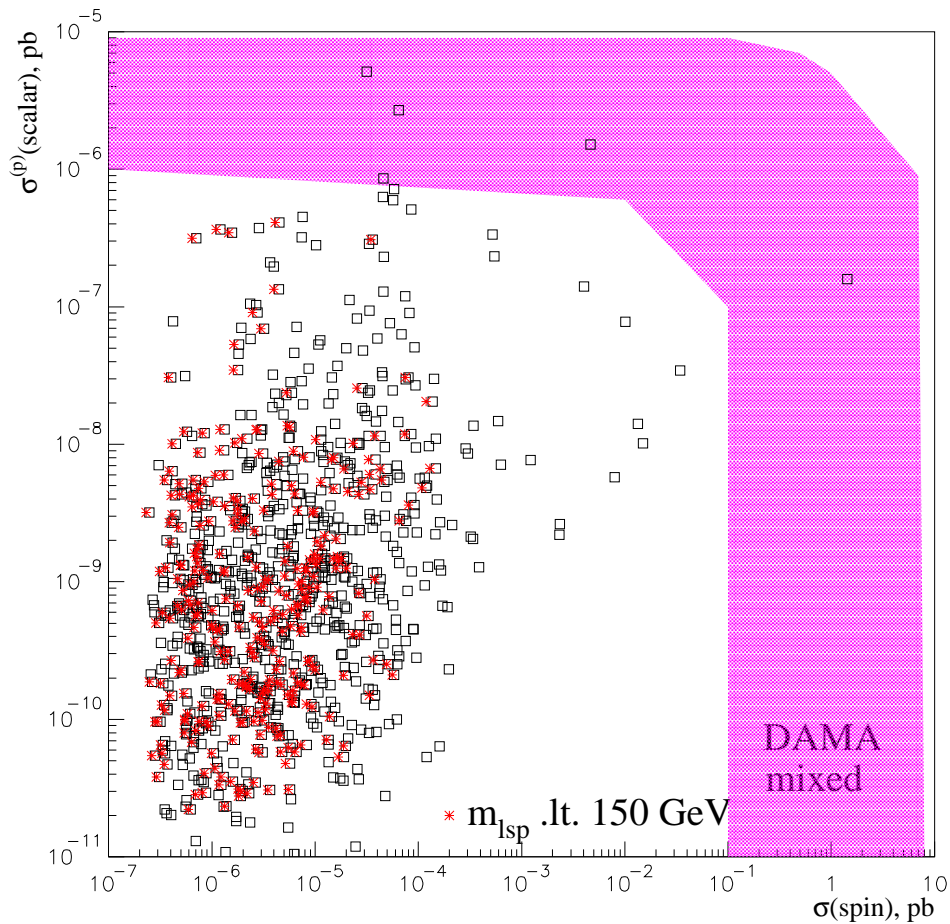}}
\end{picture}
\caption{ 
	The DAMA/NaI region 
	from the WIMP annual modulation signature (57986 kg day) 
	in the ($\xi \sigma_{\rm SI}$, $\xi \sigma_{\rm SD}$) space 
	for 40~GeV$<m^{}_{\rm WIMP}<$110~GeV covers all 
	four particular couplings ($\theta$ = 0, $\theta$ = $\pi/4$, 
	$\theta$ = $\pi/2$ and $\theta$ = 2.435 rad) reported in 
\protect\cite{Bernabei:2001ve}. 
	Scatter plots give correlations
	between $\sigma^{p}_{{\rm SI}}(0)$ and 
	$\sigma^{}_{{\rm SD}}$ in effMSSM ($\xi=1$ is assumed) 
	with the requirement for neutralino relic 
	density $~0.1< \Omega h^2 <0.3~$ and all 
	coannihilation channels included 
\protect\cite{Bednyakov:2002dz}.
	Red stars correspond to an assumption 
	that $m^{}_{\rm LSP}< 150~$GeV. }
\label{Bernabei:2001ve:fig}
\end{figure} 
	Nevertheless it was shown in 
\cite{Tovey:2000mm}
	that in the general SUSY model one can meet a case when 
	$a_{n(p)}\! \ll\! a_{p(n)}$.	
	To solve the problem (to separate SD proton and 
	neutron constraints) at least two new approaches were proposed.
	As the authors of 
\cite{Tovey:2000mm} claimed, their method 
	has the advantage that the limits on individual 
	WIMP-proton and WIMP-neutron SD cross sections 
	for a given WIMP mass can be combined 
	to give a model-independent limit 
	on the properties of WIMP scattering 
	from both protons and neutrons in the target nucleus. 
	The method relies on the assumption that the 
	WIMP-nuclear SD cross section can be presented in the form
$\displaystyle 
	\sigma^A_{\rm SD}(0) \!=\!
	\left( \sqrt{\sigma^p_{\rm SD}|^{}_A} \pm 
       \sqrt{\sigma^n_{\rm SD}|^{}_A} \right)^2$,
	where $\sigma^p_{\rm SD}|^{}_A$
	  and $\sigma^n_{\rm SD}|^{}_A$ are auxiliary 
	quantities, not directly connected with measurements.
	Furthermore, to extract, for example, a constraint on
	the sub-dominant WIMP-proton spin 
	contribution one should assume the proton contribution dominance 
	for a nucleus whose spin is almost completely 
	determined by the neutron-odd group. 
	It may seem useless,
	especially because these sub-dominant 
	constraints are always much weaker 
	than the relevant constraints 
	obtained directly with a proton-odd group target. 
	
	Another approach of  	
\cite{Bernabei:2001ve}
	is based on introduction of another auxiliary quantity 
$\displaystyle 
\sigma_{{\rm SD}} = 12\frac{\mu_{p}^2}{\pi}({a}^2_{p}+{a}^2_{n})$,
	where
$\tan \theta = {a}_{n}/{a}_{p}$.
	With these definitions the SD WIMP-proton and
	WIMP-neutron cross section can be given in the form
$\displaystyle 
\sigma^{n(p)}_{{\rm SD}}(0) = \sigma_{\rm SD}\sin^2\theta(\cos^2\theta) 
$.
	In 
Fig.~\ref{Bernabei:2001ve:fig}
	the WIMP-nucleon spin-mixed and scalar couplings 
	allowed by annual modulation signature from 
	the 100-kg DAMA/NaI experiment (57986 kg day) 
	are given by filled region. 
	The region (at 3 $\sigma$ C.L.) 
	in the ($\xi \sigma_{\rm SI}$, $\xi \sigma_{\rm SD}$) space 
	for 40~GeV$<m^{}_{\rm WIMP}<$110~GeV covers all 
	four particular couplings ($\theta$ = 0, $\theta$ = $\pi/4$, 
	$\theta$ = $\pi/2$ and $\theta$ = 2.435 rad) reported in 
\cite{Bernabei:2001ve}. 
	Scatter plots give $\sigma^{p}_{{\rm SI}}(0)$ versus
	$\sigma_{{\rm SD}}$ in effMSSM 
	with $~0.1< \Omega h^2 <0.3~$ and all 
	coannihilation channels included from 
\cite{Bednyakov:2002dz} under the assumption of $\xi=1$. 
	Red stars correspond to the same calculations
	but with $m^{}_{\rm LSP}< 150~$GeV.	
	In this mixed case the limits for the spin couplings depend on
	assumptions about the scalar coupling, 
	and the relevant exclusion curve 
	for the spin-dependent WIMP-proton cross section (not given in 
Fig.~\ref{Spin-p})	
	can not be simply extracted from these mixed results of 
\cite{Bernabei:2003xg}.

\subsection{The role of the germanium-73 isotope}
	Comparing the number of exclusion curves in 
Fig.~\ref{Spin-p} and 
Fig.~\ref{Spin-n} one can easily see that
	there are many measurements with p-odd nuclei
	and there is a lack of data for n-odd nuclei, i.e. 
	for $\sigma_{\rm SD}^{n}$. 
	Therefore measurements with  n-odd nuclei are needed.
	From our point of view this lack of $\sigma_{\rm SD}^{n}$ 
	measurements can be 
	filled with new data expected from the HDMS experiment with 
	the high-spin isotope $^{73}$Ge
\cite{Klapdor-Kleingrothaus:2002pg}.
	This isotope looks with a good accuracy 
	like an almost pure n-odd group nucleus with 
	$\langle {\bf S}_{n}\rangle\! \gg\! \langle {\bf S}_{p}\rangle$
(Table~\ref{Nuclear.spin.main.table.71-95}).
	The variation of the $\langle {\bf S}_{p}\rangle$
	and $\langle {\bf S}_{n}\rangle$
	in the table reflects the level of  
	inaccuracy and complexity 
	of the current nuclear structure calculations.
\begin{table}[h!] 
\caption{Zero-momentum spin structure (and predicted magnetic moments $\mu$) 
	of the $^{73}$Ge nucleus in different nuclear models. 
	The experimental value of the magnetic moment given in the brackets 
	is used as input in the calculations.
\label{Nuclear.spin.main.table.71-95}}
\begin{center}
\begin{tabular}{lrrr}
\hline\hline
$^{73}$Ge~($L_J=G_{9/2}$) & ~~~~~~~~$\langle {\bf S}_p \rangle$ & 
~~~~~~~~$\langle {\bf S}_n \rangle$ & ~~~~~~~~$\mu$ (in $\mu_N$) \\ \hline
ISPSM, Ellis--Flores~\cite{Ellis:1988sh,Ellis:1991ef}
	&    0	  & $0.5$		& $-1.913$ \\ 
OGM, Engel--Vogel~\cite{Engel:1989ix} 	
	&    0	  & $0.23$ 	&$(-0.879)_{\rm exp}$ \\ 
IBFM, Iachello at al.~\cite{Iachello:1991ut} and \cite{Ressell:1993qm}
	&$-0.009$ & $0.469$ &$-1.785$\\ 
IBFM (quenched), 
	Iachello at al.~\cite{Iachello:1991ut} and \cite{Ressell:1993qm}
	&$-0.005$  & $0.245$ &$(-0.879)_{\rm exp}$ \\
TFFS, Nikolaev--Klapdor-Kleingrothaus, \cite{Nikolaev:1993dd} 
	&$0$   & $0.34$ & --- \\ 
SM (small), Ressell at al.~\cite{Ressell:1993qm} 
	&$0.005$   & $0.496$ &$-1.468$ \\ 
SM (large), Ressell at al.~\cite{Ressell:1993qm} 
	&$0.011$   & $0.468$ &$-1.239$ \\ 
SM (large, quenched), Ressell at al.~\cite{Ressell:1993qm} 
	&$0.009$   & $0.372$ &$(-0.879)_{\rm exp}$ \\ 
``Hybrid'' SM, Dimitrov at al.~\cite{Dimitrov:1995gc}           
	& $0.030$ & $0.378$ & $-0.920$ \\ 
\hline\hline
\end{tabular} \end{center}
\end{table} 

\begin{figure}[!h] 
\begin{picture}(100,110)
\put(-15,-40){\includegraphics{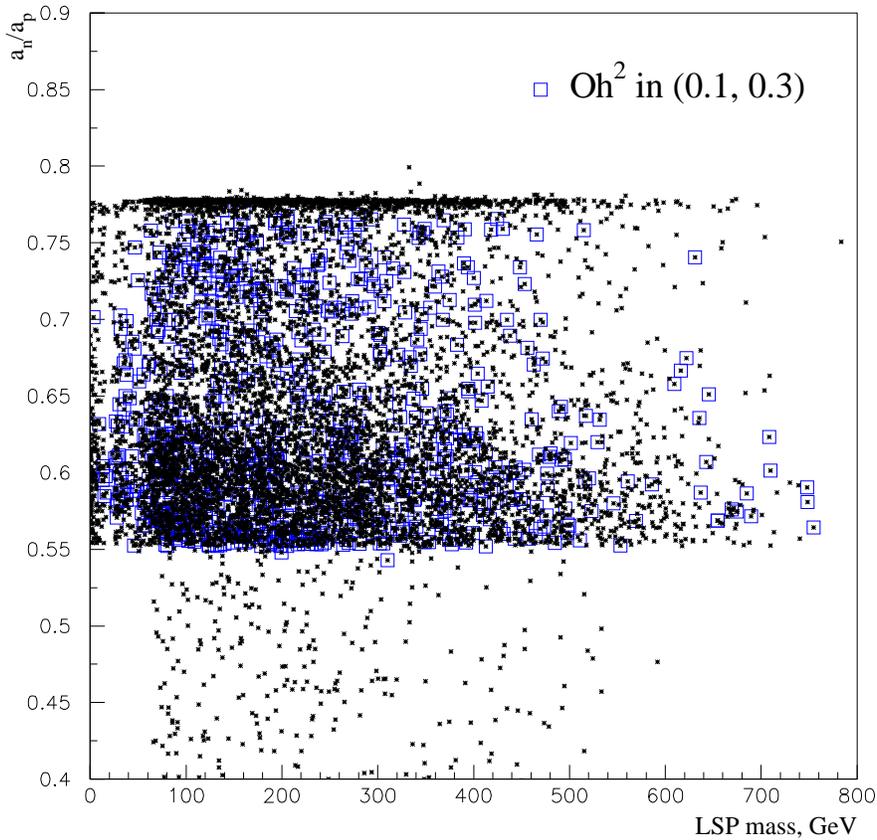}}
\end{picture} 
\caption{
	The scatter plot (dots) gives the ratio of the neutralino-neutron 
	$a_n$ and neutralino-proton $a_p$ spin couplings in the effMSSM 
\cite{Bednyakov:2002dz}.
	Boxes correspond to the relic neutralino density constraint 
	$0.1< \Omega h^2 <0.3$ in the same model.}
\label{Ratio}
\end{figure}

	On the other hand, 
Fig.~\ref{Ratio} shows that for the ratio of $a_n$ to $a_p$ 
	one can have the bounds
$$0.55 < \left|\frac{a_n}{a_p} \right| <  0.8.
$$
	The scatter plots in 
Fig.~\ref{Ratio} were obtained in effMSSM  
\cite{Bednyakov:2002dz} when all coannihilation channels were included. 
	The blue squares (black points) were calcualted 
	with (without) the relic neutralino density 
	constraint $~0.1< \Omega h^2 <0.3$. 
	Therefore in the model the couplings are almost the same
	and one can safely neglect the 
	$\langle{\bf S}^A_{p}\rangle$-spin 
	contribution in the analysis of the DM data with the $^{73}$Ge 
	target (for which 
	$\langle{\bf S}^A_{p}\rangle \ll 
	 \langle{\bf S}^A_{n}\rangle$).

	We would like to advocate the old odd-group-like approach
	for experiments with germanium detectors. 
	Of course, from measurements with $^{73}$Ge one can extract 
	not only the dominant constraint for WIMP-nucleon coupling
	$a_n$ (or $\sigma_{\rm SD}^{n}$) 
	but also the constraint for the sub-dominant WIMP-proton coupling
	$a_p$ (or $\sigma_{\rm SD}^{p}$) using the approach of 
\cite{Tovey:2000mm}.
	Nevertheless, the latter constraint will be much weaker
	in comparison with the constraints from p-odd group
	nuclear targets, like $^{19}$F or NaI.  
	The fact illustrates the NAIAD (NaI, 2003) curve in 
Fig.~\ref{Spin-n}, which corresponds to the sub-dominant
	WIMP-neutron spin contribution 
	extracted from the p-odd nucleus Na. 

\subsection{Finite Momentum Transfer}
	As $m_{\tilde \chi}$ becomes larger, 
	the product $qR$ ceases to be negligible and 
	the finite momentum transfer limit
	must be considered for heavier nuclei.  
	With the isoscalar coupling constant $a_0 = a_n + a_p$
	and the corresponding isovector coupling constant
	$a_1 = a_p - a_n$ one splits $S^A_{\rm SD}(q)$ into a pure
	isoscalar, $S_{00}$, a pure isovector, $S_{11}$, 
	and an interference term, $S_{01}$
\cite{Ressell:1997kx,Ressell:1993qm}:
\begin{equation}
\label{Definitions.spin.decomposition}
S^A_{\rm SD}(q) = a_0^2 S^A_{00}(q) + a_1^2 S^A_{11}(q) + a_0 a_1 S^A_{01}(q).
\end{equation} 
	The differential SD event rate has the form
\begin{eqnarray}
\label{Definitions.spin.differential.rate}
\frac{dR^A_{\rm SD}}{dq^2}\!
&=&\!{\rho\over{m_{\tilde \chi}m_A}} \int v dv f(v) 
		  {{8 G_F^2}\over{(2J + 1) v^2}} S^A_{\rm SD}(q).
\end{eqnarray}
	Comparing the differential rate 
(\ref{Definitions.spin.differential.rate}) 
	together with the spin structure functions of
(\ref{Definitions.spin.decomposition}) with the observed recoil spectra 
	for different targets (Ge, Xe, F, NaI, etc)
	one can directly and simultaneously 
	restrict both isoscalar and isovector 
	neutralino-nucleon effective couplings $a_{0,1}$.
	These constraints will impose most 
	model-independent restrictions on the MSSM parameter space
	free from any assumption of 
\cite{Tovey:2000mm,Bernabei:2001ve}. 
	Perhaps, it would be the best to fit all data directly 
\cite{Tovey:2000mm}
	in terms of neutralino proton and neutron effective spin couplings 
	$a^{}_{0,1}$  or $a^{}_{p,n}$
	(see for example analysis of
\cite{Miuchi:2002zp})
	and not to use such spin quantities as
	$\sigma_{\rm SD}^{p,n}$ and $\sigma^{}_{\rm SD}$.

\begin{figure}[!h] 
\begin{picture}(100,100)
\put(-35,120){\includegraphics{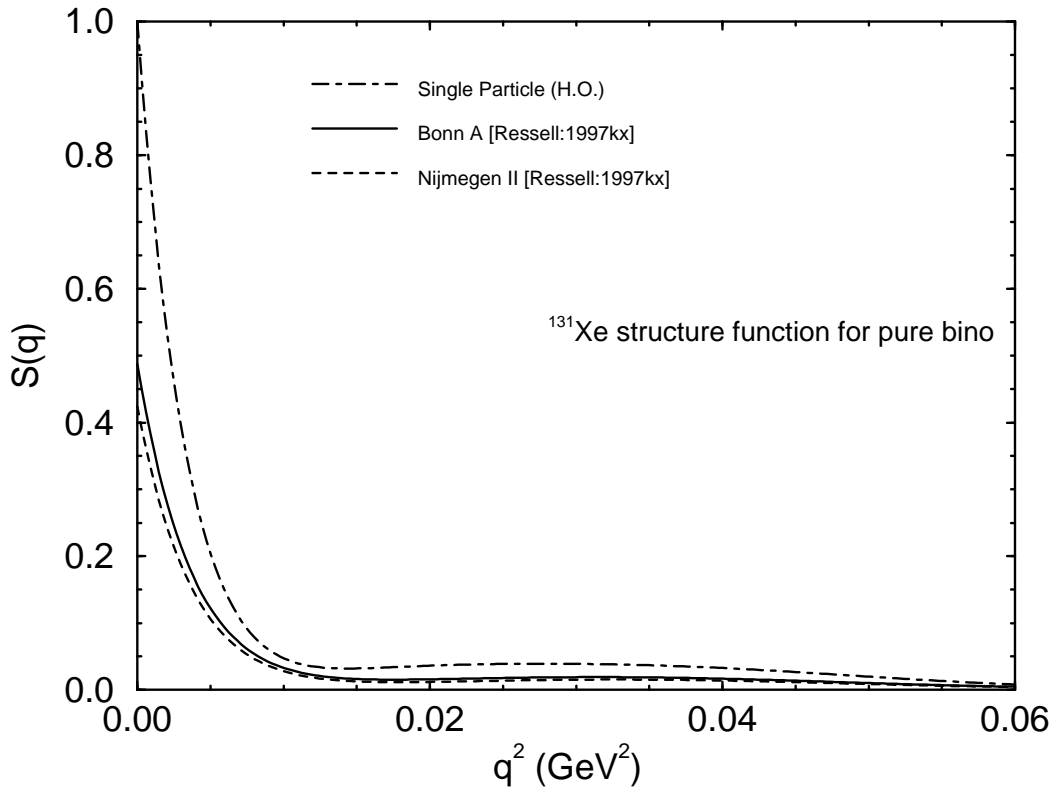}}
\end{picture} 
\caption{The $^{131}$Xe structure function for a pure bino. 
	The single-particle structure function has been
	normalized to $S(0) = 1$.  
	From 
\protect\cite{Ressell:1997kx}.
}
\label{Xe-SF}
\end{figure}

	Another attractive feature of the spin-dependent 
	WIMP-nucleus interaction is the $q$-dependence of 
	SD structure function 
(\ref{Definitions.spin.decomposition}). 
	One knows that 
	the ratio of SD to SI rate in the $^{73}$Ge detector 
	grows with the WIMP mass
\cite{Bednyakov:2000he,Bednyakov:2002mb}. 
	The growth is much greater for heavy target isotopes like xenon.
	The reason is the different behavior of 
	the spin and scalar structure functions with 
	increasing momentum transfer. 
	For example, the xenon SI structure function
	vanishes for $q^2\approx 0.02$~GeV, but 
	the SD structure function is a non-zero constant in the region
(Fig.~\ref{Xe-SF}).    

	As noted by Engel in 
\cite{Engel:1991wq},  
	the relatively long tail of the SD structure function is 
	caused by nucleons near the Fermi surface, which do the 
	bulk of the scattering. 
	The core nucleons, which dominate the SI nuclear coupling, 
	contribute much less at large $q$.
	Therefore the SD efficiency for detection of a
	DM signal is higher than the SI efficiency, 
	especially for very heavy neutralinos. 

\subsection{Conclusion} 
	The idea of this review paper is to attract 
	attention to the role of the 
	spin-dependent WIMP-nucleus interaction in the dark 
	matter search experiments. 
	The importance of this interaction is discussed.
	The fullest possible set of currently available exclusion curves 
	for spin-dependent WIMP-proton and WIMP-neutron
	cross sections is given in 
Fig.~\ref{Spin-p} and 
Fig.~\ref{Spin-n}. 
	Nowadays about two-orders-of-magnitude improvement 
	of the current DM experiment sensitivities is needed 
	to reach the SUSY predictions for the $\sigma^{p,n}_{{\rm SD}}$. 
	It is noted that a near-future experiment like HDMS
\cite{Klapdor-Kleingrothaus:2002pg}
	with the high-spin isotope 
	$^{73}$Ge being an almost pure n-odd nucleus
	can fill in this gap and can 
	supply us with new important constraints for SUSY models.

\smallskip 
	The author thanks Prof. H.V.~Klapdor-Kleingrothaus 
	for permanent collaboration and interest in the work,
	Drs. V.A.~Kuzmin and F.~Simkovic for helpful discussions  
	and the RFBR (Grant 02--02--04009) for the support.

\providecommand{\href}[2]{#2}\begingroup\raggedright\endgroup

\end{document}